\documentclass[aps,preprint,showpacs,superscriptaddress]{revtex4}
\def\comment#1{}

\usepackage{graphicx}

\begin{document}
\title{Benchmarks for the Forward Observables 
at RHIC, the Tevatron-run II and the LHC
}
\author{J.R. Cudell}
\affiliation{Institut de Physique,
B\^at. B5, Universit\'e de Li\`ege, Sart Tilman, B4000 Li\`ege, Belgium}
\author{V.V. Ezhela}
\affiliation{COMPAS group, IHEP, Protvino, Russia}
\author{P. Gauron}
\affiliation{LPNHE \cite{CNRS}-Theory Group, Universit\'e
Pierre et Marie Curie, Tour 12 E3, 4 Place Jussieu, 75252
Paris Cedex 05, France}
\author{K. Kang}
\affiliation{Physics Department, Brown University, Providence,
RI 02912, U.S.A.}
\author{Yu.V. Kuyanov}
\affiliation{COMPAS group, IHEP, Protvino, Russia}
\author{S.B.~Lugovsky}
\affiliation{COMPAS group, IHEP, Protvino, Russia}
\author{E. Martynov}
\affiliation{Institut de Physique,
B\^at. B5, Universit\'e de Li\`ege, Sart Tilman, B4000 Li\`ege, Belgium}
\affiliation{Bogolyubov Institute for Theoretical Physics, 03143
Kiev, Ukraine}
\author{B. Nicolescu}
\affiliation{LPNHE \cite{CNRS}-Theory Group, Universit\'e
Pierre et Marie Curie, Tour 12 E3, 4 Place Jussieu, 75252
Paris Cedex 05, France}
\author{E.A. Razuvaev}
\affiliation{COMPAS group, IHEP, Protvino, Russia}
\author{N.P. Tkachenko}
\affiliation{COMPAS group, IHEP, Protvino, Russia}
\collaboration{COMPETE \cite{byline} collaboration}
\date{\today}
\begin{abstract}
We present predictions on the total cross sections 
and on the ratio of the real part to the imaginary part of the 
elastic amplitude
($\rho$ parameter) for present and future $pp$ and $\bar p p$
colliders, and on total cross sections 
for $\gamma p \to$ hadrons at cosmic-ray energies and
for $\gamma\gamma\to$ hadrons up to $\sqrt{s}=1$ TeV. 
These predictions are based on an 
extensive study of possible analytic parametrisations 
invoking the biggest hadronic dataset
available at $t=0$. 
The uncertainties on total cross sections, including the systematic errors 
due to contradictory data points from FNAL,
can reach $1.9\%$ at RHIC, $3.1\%$ at the Tevatron, and $4.8\%$ at the LHC,
whereas
those on the $\rho$ parameter are respectively $5.4\%$, $5.2\%$, and $5.4\%$.
\end{abstract}
\pacs{
13.85.-t, 11.55.-m, 12.40.Nn, 13.60.Hb}

\maketitle
In recent works \cite{Cudell2002,PDB}, we have performed an exhaustive 
study of the analytic parametrisations of soft data at
$t=0$. For this purpose, we gathered the largest
available set of data at \( t=0 \), which includes all measured total cross
sections and ratios of the real part to the imaginary part of the elastic 
amplitude
(\( \rho \) parameter) for the scattering of \( pp, \) \( \overline{p}p \),
\( \pi ^{\pm }p \), \( K^{\pm }p \), and total cross sections for \( \gamma p \),
\( \gamma \gamma \) and \( \Sigma ^{-}p \) \cite{RPP2002,foot1}.
 
Several
experiments are under way \cite{RHIC}, or being planned, to measure the
hadronic amplitudes at $t=0$. Some authors \cite{Dutta:2000hh,Baur:2001jj}
also presented what they feel are
reference values for the total $\gamma p$ and $\gamma\gamma\to$ hadrons
cross sections. Thus it is timely and appropriate to present independently
our predictions for the forward observables at RHIC, the Tevatron-run II
and the LHC as well as for $\gamma p$ total cross
section at cosmic-ray energies and for $\gamma \gamma$ total cross sections
up to 1 TeV.

We can summarize the general form of the parametrisations by quoting
the form of total cross sections, from which the $\rho$ parameter is obtained
via analyticity. The ingredients are the contribution $Y^{ab}$ of 
the highest meson
trajectories ($\rho$, $\omega$, $a$ and $f$) and the rising $C=+1$
term $H^{ab}$
from the pomeron contribution to the total cross section, which can be
written for the scattering of $a$ on $b$:
\begin{equation}
\sigma_{tot}^{ab}=(Y^{ab}+H^{ab})/s
\label{mods}
\end{equation}
The first term is parametrised via Regge theory, and we allow the lower 
trajectories to be partially degenerate,
{\it i.e.} our experience shows that it is enough to introduce
one intercept
for the $C=+1$ trajectories, and another one for the $C=-1$ \cite{CKK}. 
A further lifting of the degeneracy is certainly possible, but does not 
seem to modify significantly the results \cite{Martynov2002}. Hence we 
use
\begin{equation}
Y^{ab}= Y^{ab}_{+} \left({s/s_1}\right)^{\alpha _{+}}
\pm 
Y^{ab}_{-} \left({s/s_1}\right)^{\alpha _{-}}
\label{lower}
\end{equation}
with $s_1 = 1$ GeV$^2$. The contribution of these trajectories is represented
by RR in the model abbreviations.

As for the part rising with energy, we consider here two main options:
it can rise as a $\log s$, or as a $\log^2 s$,
with in each case the possibility to add a constant term. 
We shall not consider
the simple-pole parametrisation \cite{DoLa}, 
not only because it is disfavored by
our ranking procedure (see below), but also because we want to make 
predictions at very high
energies, where unitarisation must set in \cite{foot2}. 
In the following, we shall only refer explicitly to our 
preferred parametrisation of $H^{ab}$, which we note as PL2:
\begin{eqnarray}
{{H}}^{ab}&=&s(B^{ab}\ln^{2}(s/s_0)+P^{ab})\label{pom3}
\end{eqnarray} 
where $s_0$ is
a universal scale parameter (to
be determined by the fits) identical for all collisions.

We have considered several possible constraints on the parameters
of Eqs.~(\ref{lower}-\ref{pom3}):
degeneracy of the reggeon trajectories (\( \alpha _{+}=\alpha _{-} \));
universality of rising terms ($B^{ab}$ independent of the hadrons)
\cite{universal,Gauron:2000ri};
factorization for the residues in the case of the \( \gamma \gamma \)
and \( \gamma p \) cross sections (\( H_{\gamma \gamma }=\delta H_{\gamma p}
=\delta ^{2}H_{pp} \));
quark counting rules \cite{qc} (predicting the $\Sigma p$ cross section from
$pp$, $Kp$ and $\pi p$); 
and finally the Johnson-Treiman-Freund \cite{jtf}
relation for the cross section differences.

Out of the 256 possible variants, we showed that 24 
met our criteria for applicability (an overall $\chi^2/dof\leq 1.0$ and a
non-negative pomeron
contribution at all energies) 
if one fitted
only to $\sigma_{\rm tot}$ for $\sqrt{s}\geq 10$ GeV, and 5 did for $\sqrt{s}\geq 4$ GeV (see Table XI from \cite{Cudell2002}),
whereas 20 (resp. 4) variants
obeyed this criterion when
a fit to both $\sigma_{\rm tot}$
and $\rho$ was performed, for $\sqrt{s}\geq 10$ (resp. 5) GeV (see Table XIV
from \cite{Cudell2002}).
We shall neither give here the list of models, nor spell out ranking criteria
based on new indicators that quantify certain qualities of
the fits, but simply mention that the triple-pole parametrisation
RRP$_{nf}$L2$_u$ \cite{universal,Gauron:2000ri} 
was determined to be the highest-ranking
model leading to
the most satisfactory description of the data (see similar conclusions in 
\cite{Igi}). 
This parameterization has a universal (u)
$B\log^2(s/s_0)$ term, a non-factorizing (nf) constant term and non-degenerate
lower trajectories. 

We start by giving the predictions of this
model, adjusted for ($\sqrt{s} \geq 5$ GeV), with updated data points from 
ZEUS \cite{ZEUS}. 
These predictions include statistical errors calculated from
the full error matrix $E_{ij}$. We define
\begin{equation}
\Delta Q=\sum_{ij} E_{ij} {\partial^2 Q\over\partial x_i \partial x_j}
\label{staterr}
\end{equation} 
with $Q=\sigma_{tot}$ or $\rho$ and $x_i$ the parameters of the model. 
These errors are shown in Figs.~1 and 2 by a filled band, and
in Tables \ref{table2}, \ref{table3}, and \ref{table4}.
\begin{figure}
\includegraphics[scale=0.5]{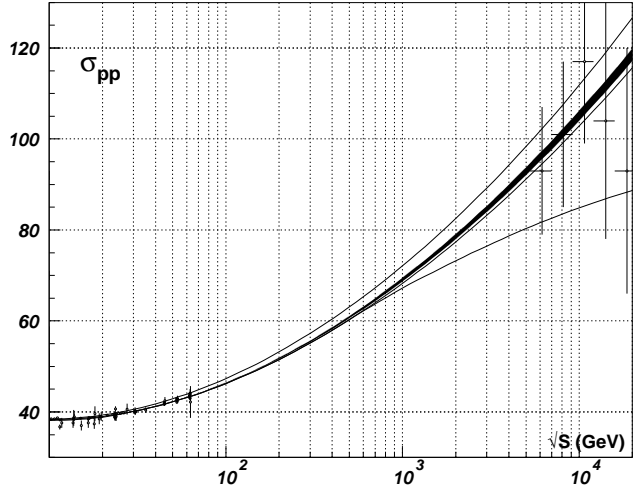}
\caption{
Predictions for total cross sections. The black
error band shows the statistical errors to the best fit,
the closest curves near it give the sum of statistical and systematic
errors to the best fit due to the ambiguity in Tevatron data
, and the highest
and lowest curves show the total errors bands
from all models considered in this letter
(note that the upper curve showing the systematic error is 
indistinguishable from the highest curve in this case).
}
\label{sigtot}
\end{figure}
\begin{figure}
\includegraphics[scale=0.5]{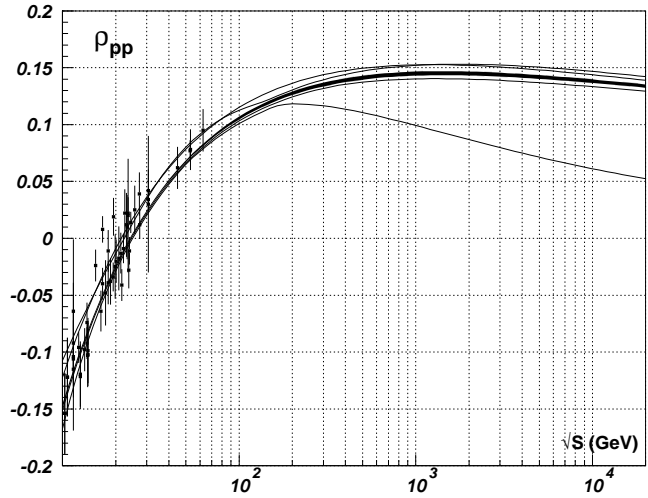}
\caption{Predictions for the $\rho$ parameter.
The curves and band
are as in Fig.~1.
}
\label{rho}
\end{figure}

In these figures and tables, we also give our estimate of the 
systematic uncertainty
coming from the discrepancy between different FNAL measurements of 
$\sigma_{tot}$: we fit RRP$_{nf}$L2$_u$ either to the high data (CDF) or to 
the low ones (E710/E811), and get two error bands. The distances from the
central value of the combined fit 
to the upper (resp. lower) border of these bands give us 
the positive
(resp. negative) systematic errors. 
We estimate the total errors as the sum of the systematic and of the statistical
uncertainties \cite{footerr}. 

One can see that the total errors on total cross sections are of the order
of 1.9$\%$ at RHIC, of the order of $3.1\%$ at the Tevatron and as large
as $4.8\%$ at the LHC and dominated by the systematic errors.
The errors on the $\rho$ parameter are much larger, reaching 5.4$\%$ at RHIC,
5.2$\%$ at the Tevatron and 5.4$\%$ at the LHC. 
This is due to the fact that experimental errors are bigger, 
hence less constraining, but this also stems
from the incompatibility of some low-energy determinations of $\rho$ 
\cite{Cudell2002}. This means that the systematic error is always 
bigger than the statistical one. 

Concerning the contradictory data, we are forced to use them in our fits
until the discrepancy is resolved by further experiments. 
In the case of the Tevatron data, one can see that the discrepancy
results in a big shift (of more than
$1\sigma$) in the central value of the coefficient $B$ of the $\log^2 s$ term, 
which controls the asymptotic behavior, and hence that asymptotic
predictions are appreciably weakened by the present situation. 
The opportunities to measure $\sigma_{tot}$ and $\rho$ will
be scarce in the future, hence
any new measurement 
at RHIC, the Tevatron run II and the LHC should not be missed. Unfortunately,
the recent publication of E-811 \cite{Avila:2002bp} does not clear the 
problem as their value
for $\rho$ is fully compatible with our preferred model, whereas their 
number for $\sigma_{tot}$ (which is highly correlated with $\rho$)
has hardly changed.  

It is interesting to note that the choice of one FNAL result or the others
leads to 
a variation of the overall fit quality, as shown
in Table \ref{table1} (last two columns)
\cite{foot3}:
the variant with CDF data has slightly better
overall $\chi^2/dof$ and better
$\chi^2/nop$ distribution over sub-samples.
We can consider this as an indication
that the global picture emerging from fits to all data on forward observables
supports 
the CDF data and disfavors the 
E710/E811 data at $\sqrt{s}=1.8$ TeV
(see also an analogous conclusion based on the other arguments in
 \cite{Albrow:1999va}). 
\begin{table}
\caption{Summary of the quality of the fits at 
different stages of the Review of Particle Physics (RPP)  
database (DB): DB02 -- The 2002 RPP DB; DB02Z
-- the 2002 RPP DB with new ZEUS data, DB02Z-CDF -- with 
the CDF point removed;
DB02Z-E710/E811 with E710/E811 points
removed. The first line gives the overall $\chi^2/dof$ for the
global fits, the other lines give the $\chi^2/nop$ for data sub-samples,
the last line gives in each case the parameter controlling the asymptotic
form of cross sections.}
\begin{tabular}{lcccc}
\hline
 & DB02 &DB02Z &DB02Z &DB02Z \\
Sample & & &$-$CDF &$-$E710/E811\\
\hline
total & 0.968 & 0.966 & 0.964 & 0.951 \\ 
\hline
\multicolumn{5}{c}{total cross sections}\\
\hline
$\overline{p} p$ & 1.15 & 1.15 & 1.12 & 1.05 \\
$p p$ & 0.84 & 0.84 & 0.84 & 0.84 \\ 
$\pi^- p$ & 0.96 & 0.96 & 0.96 & 0.96 \\
$\pi^+ p$ & 0.71 & 0.71 & 0.71 & 0.71 \\
$K^- p$ & 0.62 & 0.62 & 0.62 & 0.61 \\
$K^+ p$ & 0.71 & 0.71 & 0.71 & 0.71 \\
$\Sigma^- p$ & 0.38 & 0.38 & 0.38 & 0.38 \\
$\gamma p$ & 0.61 & 0.58 & 0.58 & 0.58 \\
$\gamma \gamma$ & 0.65 & 0.64 & 0.64 & 0.63 \\
\hline
\multicolumn{5}{c}{elastic forward Re/Im}\\
\hline
$\overline{p} p$ & 0.52 & 0.52 & 0.52 & 0.53 \\
$p p$ & 1.83 & 1.83 & 1.83 & 1.80 \\ 
$\pi^- p$ & 1.10 & 1.10 & 1.09 & 1.14 \\
$\pi^+ p$ & 1.50 & 1.50 & 1.52 & 1.46 \\
$K^- p$ & 1.00 & 0.99 & 1.01 & 0.96 \\
$K^+ p$ & 1.07 & 1.07 & 1.10 & 0.98 \\
\hline
\multicolumn{5}{c}{ values of the parameter B }\\
\hline
 &0.307(10)&0.307(10)&0.301(10)&0.327(10)\\
\hline 
\end{tabular}
\label{table1}
 \end{table}

Finally, we also present in Figs.~1 and 2 our estimate of the region
where new physics would be discovered. 
For each of the 20 parametrisations which satisfy
our criteria for applicability \cite{Cudell2002}
for $\sqrt{s}\geq 10$ GeV, and which obey the Froissart-Martin bound 
\cite{foot2,bound}, 
we construct errors bands according to Eq.~(\ref{staterr}). 
This gives us 20 $1\sigma$-error bands. Their union
represents the ``allowed region" where analytic models built according
to (\ref{mods}) can reproduce the data.
A measurement outside of this region would imply that 
new physics ingredients are needed. 

To conclude, we believe that we have given here the best possible estimates
for present and future $pp$ and $\bar p p$ facilities. Although one might
be tempted to use only data in an energy range close to the one measured,
one must realize that analytic parametrisations are constrained both by
lower-energy data, and by their asymptotic regime. Because the pomeron
mixes (physically and numerically) with the $f$ trajectory, fits to 
all data help to disentangle the two contributions. 

A sharpening of our error bars would enable one to decide if the
unitarisation plays an essential role
and what form it takes.
This in turn can have an impact on the determination of 
the survival of
probability gaps in hard scattering, and on the usefulness of pomeron
exchange as a detection tool. 

Any significant deviation from the predictions based on model RRP$_{nf}$L2$_u$ 
will lead to a re-evaluation of the hierarchy of models and presumably
change the preferred parametrisation to another one. 
A deviation from the ``allowed region" 
would be an indication that 
strong interactions demand
a generalization of the analytic models discussed so far, {\it e.g.} by adding
odderon terms, or new pomeron terms, as suggested by QCD. 
\begin{table}
\caption{Predictions for $\sigma_{tot}$ and $\rho$, for $\bar pp$ (at
$\sqrt{s}=1960$ GeV) and for $pp$ (all other energies). The central values 
and statistical errors correspond to the preferred model RRP$_{nf}$L2$_u$, 
and the systematic errors come from the consideration of two choices between
CDF and E-710/E-811 ${\overline p} p$ data in the simultaneous global fits.}
\begin{tabular}{ccc}
\( \sqrt{s} \) (GeV)& \( \sigma \) (mb)& \( \rho \)\\
\hline
100& \( 46.37\pm 0.06\begin{array}{c} +0.17\\ -0.09 \end{array} \)& 
 \( 0.1058\pm 0.0012\begin{array}{c} +0.0040\\ -0.0021 \end{array}\)\\
200& \( 51.76\pm 0.12\begin{array}{c} +0.39\\ -0.21 \end{array} \)& 
 \( 0.1275\pm 0.0015\begin{array}{c} +0.0051\\ -0.0026 \end{array}\)\\
300& \( 55.50\pm 0.17\begin{array}{c} +0.57\\ -0.30 \end{array} \)&
 \( 0.1352\pm 0.0016\begin{array}{c} +0.0055\\ -0.0028 \end{array}\)\\
400& \( 58.41\pm 0.21\begin{array}{c} +0.71\\ -0.36 \end{array} \)&
 \( 0.1391\pm 0.0017\begin{array}{c} +0.0056\\ -0.0030 \end{array}\)\\
500& \( 60.82\pm 0.25\begin{array}{c} +0.82\\ -0.45 \end{array} \)& 
 \( 0.1413\pm 0.0017\begin{array}{c} +0.0057\\ -0.0030 \end{array}\)\\
600& \( 62.87\pm 0.28\begin{array}{c} +0.94\\ -0.48 \end{array} \)&
 \( 0.1416\pm 0.0018\begin{array}{c} +0.0058\\ -0.0031 \end{array}\)\\
1960& \( 78.27\pm 0.55\begin{array}{c} +1.85\\ -0.96 \end{array} \)&
 \( 0.1450\pm 0.0018\begin{array}{c} +0.0057\\ -0.0030 \end{array}\)\\
10000& \( 105.1\pm 1.1\begin{array}{c} +3.6\\ -1.9 \end{array} \)&
 \( 0.1382\pm 0.0016\begin{array}{c} +0.0047\\ -0.0027 \end{array}\)\\
12000& \( 108.5\pm 1.2\begin{array}{c} +3.8\\ -2.0 \end{array} \)& 
 \( 0.1371\pm 0.0015\begin{array}{c} +0.0046\\ -0.0026 \end{array}\)\\
14000& \( 111.5\pm 1.2\begin{array}{c} +4.1\\ -2.1 \end{array} \)& 
 \( 0.1361\pm 0.0015\begin{array}{c} +0.0058\\ -0.0025 \end{array}\)\\
\end{tabular}
\label{table2}
 \end{table}
 \begin{table}
\caption{Predictions for $\sigma_{tot}$ for $\gamma p \to hadrons$ for 
cosmic ray photons. The central values, the statistical errors and 
the systematic errors are as in Table \ref{table2}.}
\begin{tabular}{cc}
\( p_{lab}^{\gamma} \) (GeV)& \( \sigma \) (mb)\\
\hline
$0.5\cdot10^6$&\( 0.243\pm 0.009\begin{array}{c}+0.011\\ -0.010\end{array}\)\\
$1.0\cdot10^6$&\( 0.262\pm 0.010\begin{array}{c}+0.013\\ -0.011\end{array}\)\\
$0.5\cdot10^7$&\( 0.311\pm 0.014\begin{array}{c}+0.019\\ -0.015\end{array}\)\\
$1.0\cdot10^7$&\( 0.333\pm 0.016\begin{array}{c}+0.021\\ -0.017\end{array}\)\\
$1.0\cdot10^8$&\( 0.418\pm 0.022\begin{array}{c}+0.030\\ -0.024\end{array}\)\\
$1.0\cdot10^9$&\( 0.516\pm 0.029\begin{array}{c}+0.042\\ -0.032\end{array}\)\\
\end{tabular}
\label{table3}
\end{table}
\begin{table}
\caption{Predictions for $\sigma_{tot}$ for $\gamma \gamma \to hadrons$. 
The central values, the statistical errors and the systematic errors are 
as in Table \ref{table2}.}
\begin{tabular}{cc}
\( \sqrt{s} \) (GeV)& \( \sigma \) ($\mu$ b)\\
\hline
200&\( 0.546\pm 0.027\begin{array}{c}+0.027\\ -0.027\end{array} \)\\
300&\( 0.610\pm 0.035\begin{array}{c}+0.037\\ -0.035\end{array} \)\\
400&\( 0.659\pm 0.042\begin{array}{c}+0.044\\ -0.042\end{array} \)\\
500&\( 0.700\pm 0.047\begin{array}{c}+0.050\\ -0.048\end{array} \)\\
1000&\( 0.840\pm 0.067\begin{array}{c}+0.073\\ -0.069\end{array} \)\\
\end{tabular}
\label{table4}
\end{table}
\section*{Acknowledgments }
The COMPAS group was supported in part by the Russian Foundation for
Basic Research grants RFBR-98-07-90381 and RFBR-01-07-90392. K.K.
is in part supported by the U.S. D.o.E. Contract DE-FG-02-91ER40688-Task
A. We thank Professor Jean-Eudes Augustin for the hospitality at
LPNHE-Universit\'e Paris 6, where part of this work was done.
We thank \break O.~Selyugin for his corrections to the database, A.~Sobol for
discussions on the soft physics program of the TOTEM \&CMS, and
Yu.Kharlov for discussions on possibilities for
$\gamma \gamma$ physics at RHIC and at the LHC.


\begin{thebibliography}{99}
\bibitem[b]{foot2}
Note that we checked \cite{PDB} that the predictions of simple-pole models
nevertheless fall into our allowed region.
\bibitem[c]{footerr}Note that, apart from the tiny statistical error, none of
the errors or bands given in this work has a probabilistic interpretation.
\end{thebibliography}
\end{document}